\documentclass[multphys,vecphys]{svmult}

\usepackage{makeidx}         
\usepackage{graphicx}        
\usepackage{multicol}        
\usepackage[bottom]{footmisc}

\newcommand{\ddrv}[2]{\frac{\partial^{2}  #1}{{\partial #2}^2}   }

\makeindex             

\begin{document}

\title*{QPOs: Einstein's gravity non-linear resonances} 
\author{Paola Rebusco\inst{1} \and Marek A. Abramowicz\inst{2}\inst{3}}
\institute{
Max-Planck-Institut f\"ur Astrophysik, Karl-Schwarzschild-Strasse 1, 85741
Garching b.M\"unchen, Germany
\texttt{pao@mpa-garching.mpg.de}\and
 Department of Physics, G\"oteborg University, S 412 96
G\"oteborg, Sweden
 \texttt{number44@fy.chalmers.se}
\and
 N. Copernicus Astronomical Center,
Polish Academy of Sciences, Warsaw, Poland}
%
%
\maketitle

\begin{abstract}
There is strong evidence that the observed kHz Quasi Periodic Oscillations (QPOs)  in the X-ray flux of neutron star and black hole sources in LMXRBs are linked to Einstein's General Relativity. 
Abramowicz\&Klu\'zniak (2001) suggested a non-linear resonance model to explain the QPOs origin:  
here we summarize their idea and the development of a mathematical toy-model which begins to throw light on the nature of Einstein's gravity non-linear oscillations.
 
\end{abstract}

\section{Introduction}
QPOs are highly coherent Lorentzian peaks observed in the X-ray power spectra of compact objects in LMXRBs. They cover a wide range of frequencies (from mHz to kHz) and  they show up alone, in pairs or in triplets.
In the present review we focus on kHz QPOs which occur in pairs (the so called twin peak QPOs).
Many models have been proposed in order to understand kHz QPOs: some of them involve orbital motions
(e.g. \cite{Ste98}, \cite{Lam03}, but there are also models that are based on accretion disk oscillations (e.g. \cite{Wag01}, \cite{Kat01},\cite{Rez03},\cite{Lin04}).
 Here we present
semi-analytical results, connected in particular with the Klu\'zniak\&Abramowicz model.

\section{QPOs and General Relativity}
High frequency (kHz) QPOs lie in the range of orbital frequencies
of circular geodesics just few Schwarzschild radii outside the central source. Moreover 
 the frequencies scale with $1/M$, where $M$
is the mass of the central object \cite{Mcc04}.
These two facts support the strong gravity orbital oscillations models.\\
Consider a test particle rotating around a compact source: the radial epicyclic frequency of planar motion and the vertical epyciclic
 frequency of nearly off-plane motion are respectively defined as

\begin{equation}
\omega_r^2 =\left (\frac{1}{2 g_{rr}} \ddrv{U_{eff}}{r}\right )_{\!\ell,r_0,\pi/2}\:\:\:,
\quad
\omega_{z}^2 = 
\left (\frac{1}{2 g_{\theta\theta}} \ddrv{U_{eff}}{\theta}\right )_{\!\ell,r_0,\pi/2}\:\:\:,\nonumber
\end{equation}

where $U_{eff} = g^{tt} +l g^{t \phi}+l^2 g^{\phi \phi}$ is the effective potential and $(r,\theta,\phi)$ are spherical coordinates.
The eigenfrequencies depend only on the metric of the system, hence on strong gravity itself.
In Newtonian gravity there is degeneracy between these eigenfrequencies and
 the Keplerian frequency (all three
frequencies are equal) while in General Relativity this degeneracy is broken and as a consequence two 
   or three different characteristic frequencies are present, opening the possibility 
   of internal resonances .
As a consequence, while   Newtonian orbits are all close,
in GR they do not close after one
loop (this is for the same  reason as the well-known advance of the perihelion of Mercury).\\
For a spherically symmetric gravitating fluid body (a better model of the accretion
  disk) these are the frequencies at which the center of mass 
  (initially on a circular geodesic) oscillates.

\section{Klu\'zniak-Abramowicz resonance model}
 QPOs often occur in pairs, and  the centroid frequencies of these
pairs are in rational ratio (e.g.\cite{Str01}): these features suggested  that
high frequency QPOs are a phenomenon due to non-linear resonance, and that there may be  an analogy between the radial
and vertical oscillations in a Shakura-Sunyaev disk and the motion of a pendulum with oscillating point of
 suspension (\cite{AK01}). 
Since in GR $\omega_r < \omega_z$ the
first allowed resonance would appear for $\omega_z : \omega_r=3:2$ and it would be the strongest. \\
In all four microquasars which exhibit double peaks, the ratio of the two frequencies is $3:2$,
as well as in many neutron star sources. Moreover combinations of frequencies and subharmonics 
have been detected: these are all signatures of non-linear resonance and they confirm the validity of the model. 
\subsection{Toy model}

 A mathematical toy-model for the Klu\'zniak-Abramowicz resonance idea for QPOs was recently developed \cite{Abr05}.
   It describes the QPOs phenomenon in terms of two coupled non-linear forced oscillators,
  
     \begin{equation}
      \label{eqns}
   \ddot\delta r+\omega_r^2 \delta r = F(\delta r,\delta z, \dot\delta r, \dot \delta z)+C \cos{(\omega_0 t)}+N_r(t),
\end{equation}
\begin{equation}
       \ddot\delta z+\omega_{z}^2 \delta z =  G(\delta r,\delta z, \dot\delta r, \dot \delta z)+D \cos{(\omega_0 t)}+N_z(t),
      \end{equation}

where $F$ and $G$ are polynomes of second  or higher degree (obtained in terms of expansion of
 the deviations from a Keplerian flow). 
 The $\cos{(\omega_0 t)}$ terms 
  represent an external forcing: they are mostly important in the case of NS, where $\omega_0$
   can be the NS spin frequency. $N_r$ and $N_z$ describe the stochastic noise due to the Magneto-Rotational Instability .\\
In absence of turbulence, the first finding, which can be derived by using any  perturbative method, is that
in the approximate solution  there are terms with the denominators
in the form $n \omega_r-m \omega_z$ (with $n$ and $m$ being integers).
These $n$ and $m$ cannot take any value, but they depend on the symmetry
 of the metric and of the perturbation: for a plane symmetric configuration one
  can demonstrate (e.g.\cite{Reb04},\cite{Hor04}) that $m=2\:p$ ($p$ integer).
  Due to this, the regions where $n \omega_r= 2 p \omega_z$ are candidate regions
   of internal resonance , and
 the strongest one would be for $\omega_z:\omega_r=3:2$ (a similar reasoning can be done
 for the forcing term).

\subsection{Correlation and anticorrelation}
\begin{figure}
\begin{minipage}[h]{5.4 cm}
\centering
\includegraphics[height=3.8 cm]{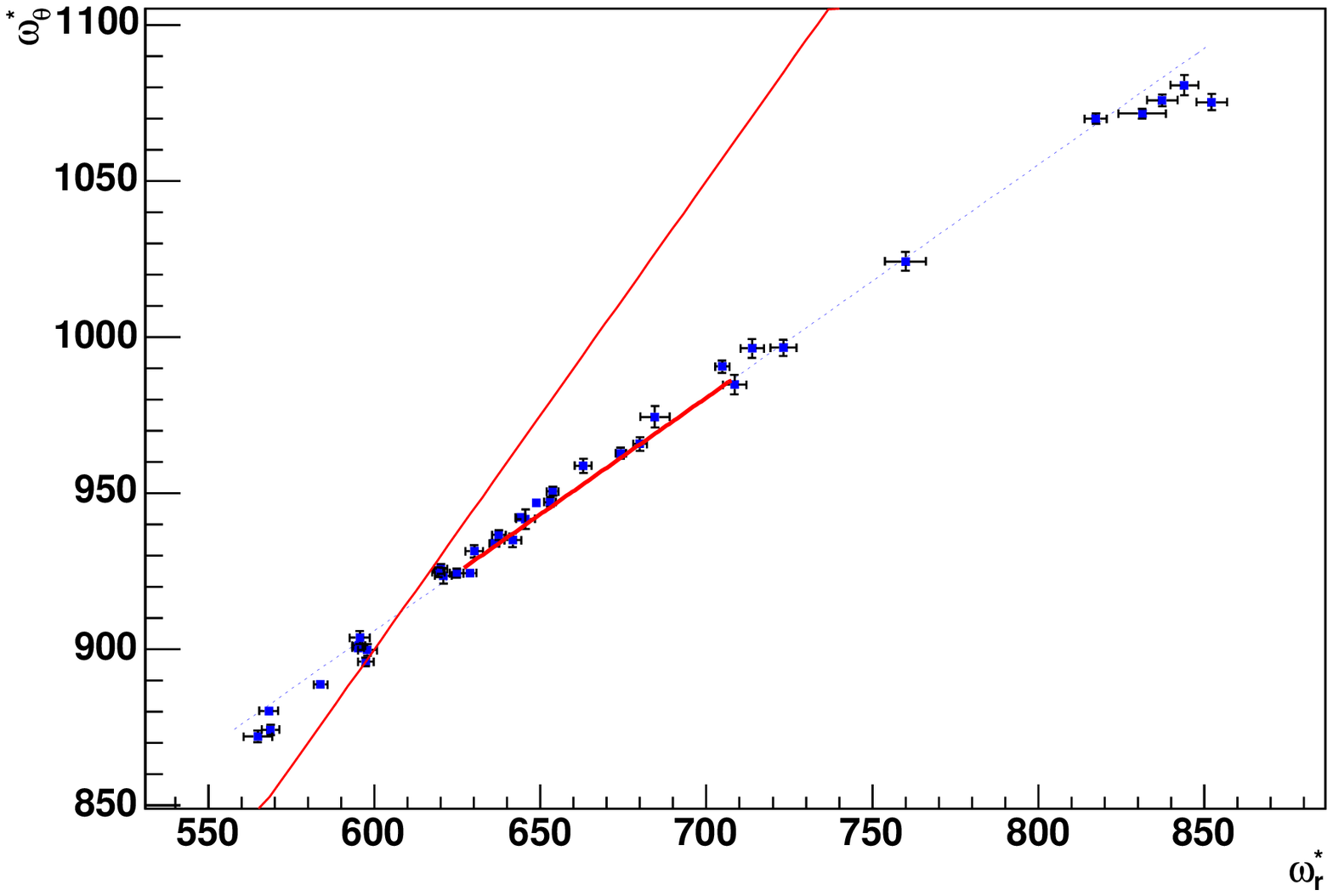}
%
%
\caption{The dotted line is the least-squares best-fit to the data
points (the  observed kHz QPOs  frequencies in Sco X-1); the thin solid line corresponds to a slope
of $3:2$ (for reference) . The thick solid segment is the analytic approximation, in which
 the frequencies are scaled for comparison with observations.}
\label{reb:fig1}       
\end{minipage}
\begin{minipage}[h]{5.4 cm}
\includegraphics[height=3.8 cm]{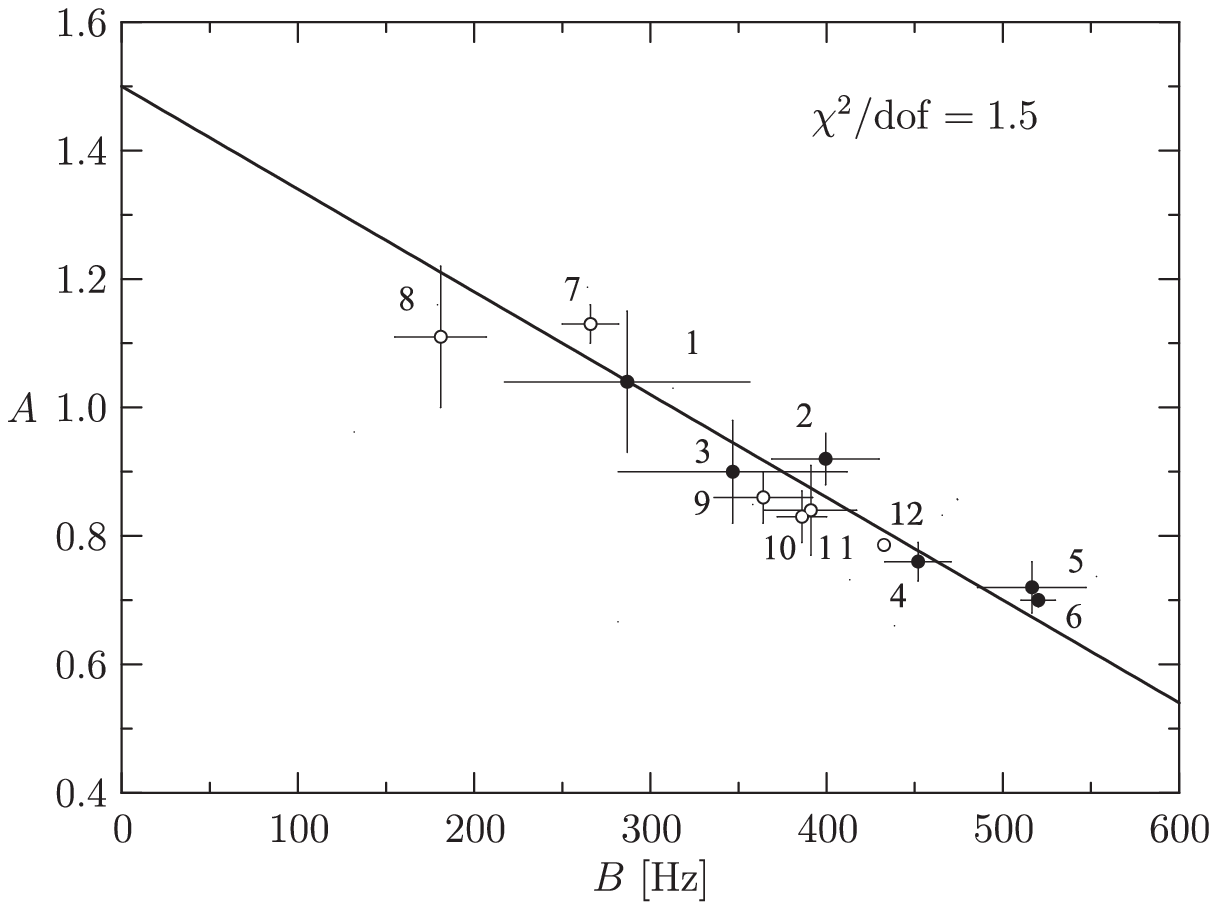}
%
%
\caption{The anticorrelation between shifts ($B$) and slopes ($A$). The points
 correspond to the individual Z and atoll sources listed in \cite{Abr05b}. The best fit line goes
 through the point $(0,1.5)$ (Courtesy of Gabriel T\"or\"ok).}
\label{reb:fig2}       
\end{minipage}
\end{figure}

A well known property of weakly non-linear oscillators is the fact that the frequencies of oscillation
($\omega_r^*$ and $\omega_z^*$)
 depend on the amplitudes, even if their value remains close to that of the eigenfrequencies ( $\omega_r$ 
and $ \omega_z$).
Perturbative methods help us again (e.g. \cite{Moo76}): indeed one can write
    
\begin{eqnarray}
     \omega_r^*&=&\omega_r+\epsilon^2 \omega_r^{(2)}+\epsilon^3 \omega_r^{(3)}+O(\epsilon^4)\:\:\:,\nonumber\\ 
     \omega_z^*&=&\omega_z+\epsilon^2 \omega_z^{(2)}+\epsilon^3 \omega_z^{(3)}+O(\epsilon^4) \end{eqnarray}

and find the frequencies corrections ($\omega_i^{(j)}$) by constraining the solution to be stationary
 at a given timescale (of the order of $\epsilon^{-j}$). For small perturbations one would get that
$\omega_z^*= A \omega_r^* + B$.
In this way one can qualitatively explain the observed linear correlation between the twin frequencies in NS sources: 
the example of the NS source Sco X-1 was studied by \cite{Abetal03} and\cite{Reb04} (see Fig.\ref{reb:fig1}).


A direct consequence of the internal resonance is the fact that  for different sources the coefficients $A$ and $B$ should be nearly linearly anticorrelated (\cite{Abr05}): $A = \omega_z/\omega_r - B/f_0(M)$, where
 $f_0(M)$ is a function which depends on the mass of the central object. Note that at higher orders 
 the relation is expected to slightly deviate from linearity: anyway up to now this feature fits very well
 the available data (see Fig.\ref{reb:fig2}).

\subsection{The effect of turbulence}
\begin{figure}
\begin{minipage}[h]{5.4 cm}
\centering
\includegraphics[height=3.8 cm]{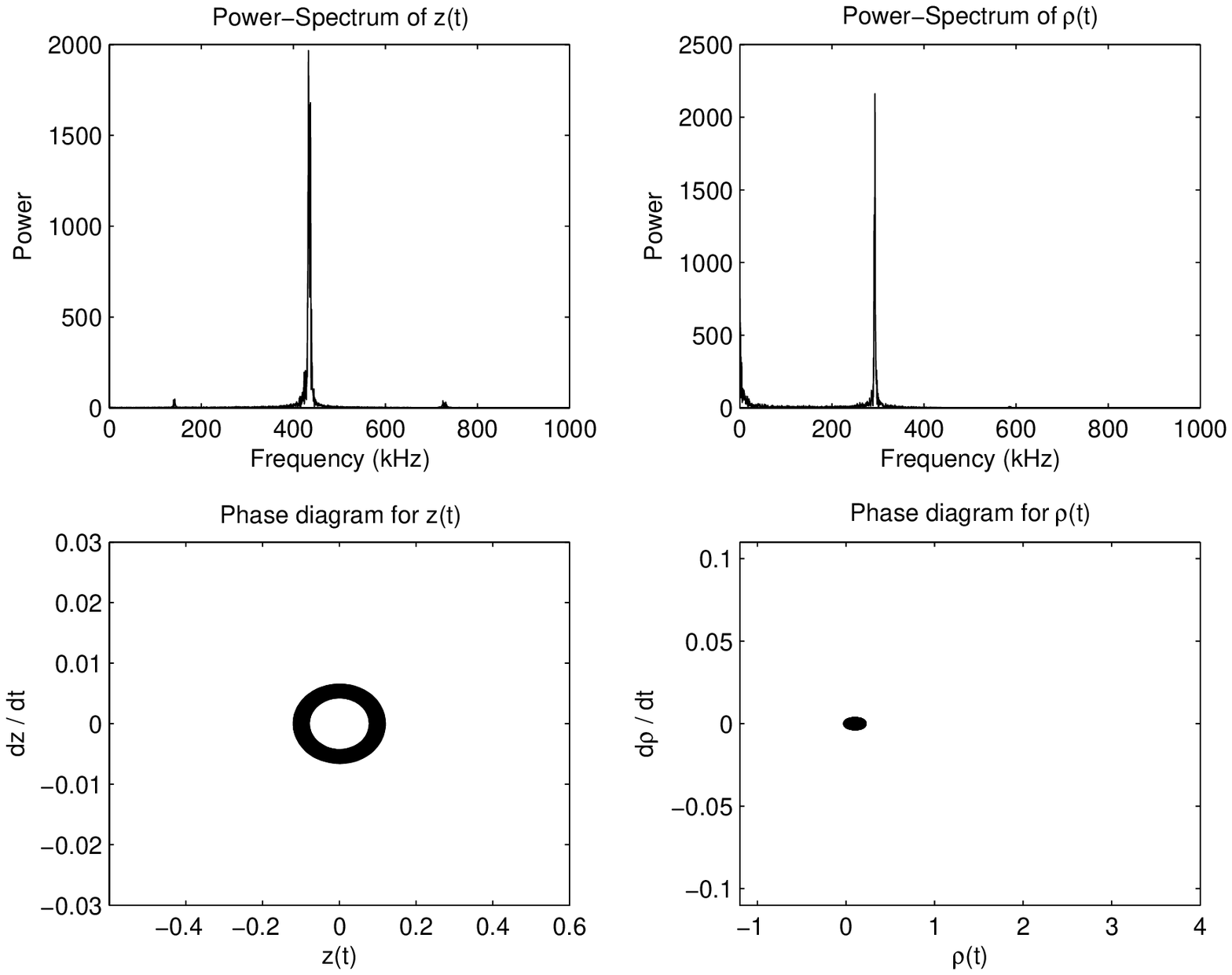}
%
%
\caption{Low turbulence: power spectra (upper part) and phase diagrams (lower part) for
	 $r$ and $z$. The displacements are in units of $r_g$, the frequencies are scaled to kHz (e.g. assuming a central mass $M$ of $2 M_\odot$). It does not differ much from the behavior in absence of turbulence.}
\label{reb:fig3}       
\end{minipage}
\hfil
\begin{minipage}[h]{5.4 cm}
\includegraphics[height=3.8 cm]{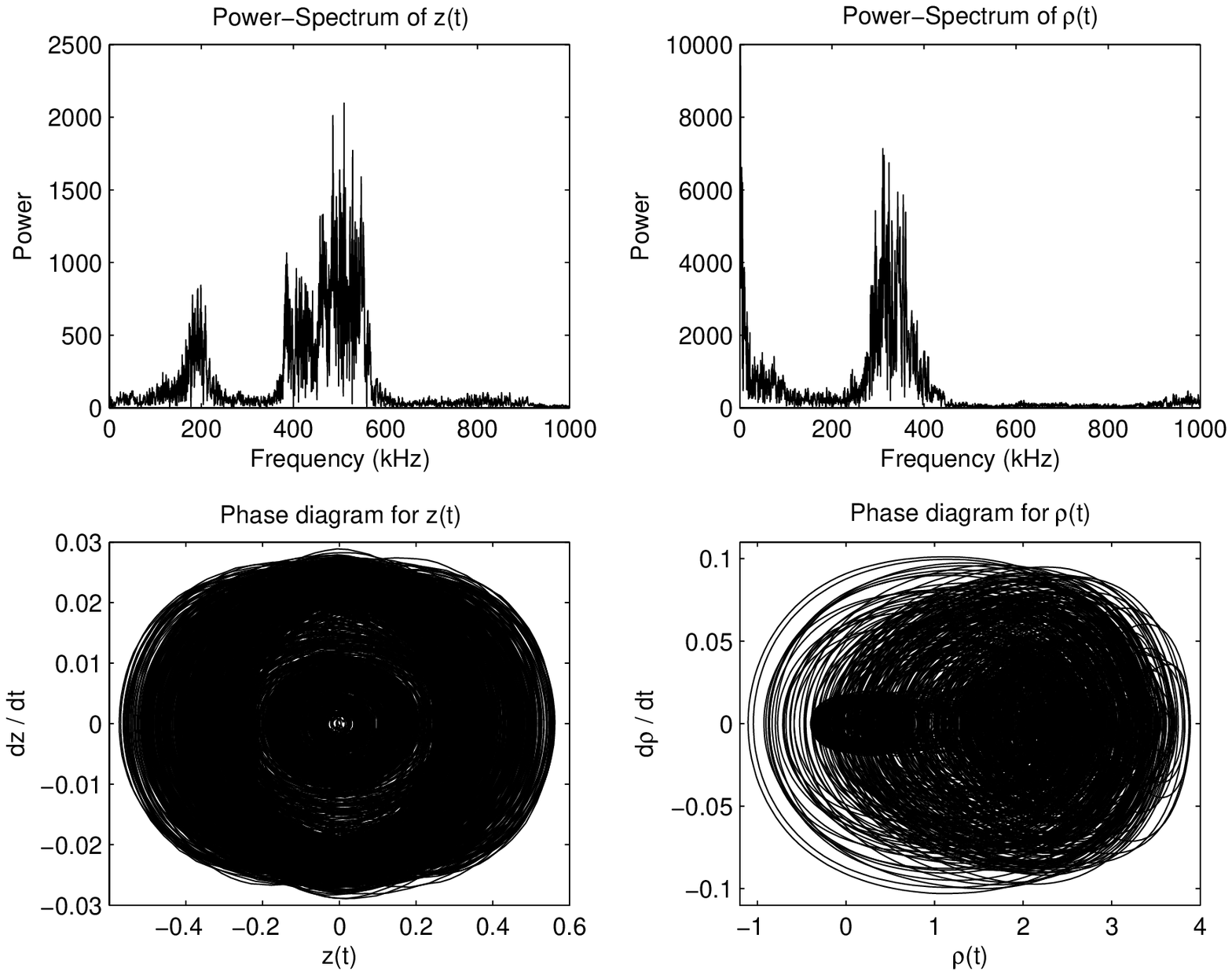}
%
%
\caption{The same as in the previous figure. 
In this case however the turbulence is strong enough to feed
 the resonance: as
 a consequence the amplitudes of oscillation are much greater.}
\label{reb:fig4}       
\end{minipage}
\end{figure}

Accretion disks  are characterized by a huge number of degrees of
freedom. The turbulent processes  can be assumed to have a stochastic nature. In particular, we have 
investigated a simplified model for the Klu\'zniak-Abramowicz nonlinear theory and showed that 
a small noise in the vertical direction can trigger coupled epicyclic oscillations (see Fig.\ref{reb:fig3}
 and \ref{reb:fig4}).
On the other hand too much noise would disrupt the quasi-periodic motion  \cite{Vioetal05}.
This is similar to the stochastically excited p-modes in the Sun, and it may help in estimating
 the strength and the nature of the turbulence itself.


\newpage

\section{Conclusions}
Non-linear parametric resonances occur everywhere in Nature: together with GR they could  explain the mechanism at the basis of kHz QPOs. 
In this way the mass and the angular momentum (e.g.\cite{Asc04}) of the central compact object could be precisely measured, but most of all Einstein's strong gravity could be proved: a good motivation to keep on investigating on this puzzling phenomenon.


\printindex
\end{document}